\documentclass[pra,a4paper,showpacs,superscriptaddress]{revtex4-2}
\usepackage{amsmath,amscd,amsfonts,amssymb, amsbsy, color,bbm, bm}
\usepackage{graphicx}
\usepackage{enumitem}
\newcommand{\be}{\begin{equation}}
\newcommand{\ee}{\end{equation}}

\newcommand{\ba}[1]{\left(\begin{array}{#1}}
\newcommand{\ea}{\end{array}\right)}

\numberwithin{equation}{section}
\begin{document}
\title{Geometric picture for SLOCC classification of \\  pure permutation symmetric three-qubit states} 
\author{K. Anjali} 
\affiliation{Department of Physics, Bangalore University, Bangalore-560 056, India} 

\author{I. Reena} 
\affiliation{Department of Physics, Bangalore University, Bangalore-560 056, India}

\author{Sudha*} 
\affiliation{Department of Physics, Kuvempu University, 
	Shankaraghatta-577 451, Karnataka, India}
\affiliation{Inspire Institute Inc., Alexandria, Virginia, 22303, USA.}
\email{tthdrs@gmail.com} 

\author{B.G.Divyamani} 
\affiliation{Department of Physics, Tunga Mahavidyalaya, Thirthahalli-577432, India}

\author{H. S. Karthik} 
\affiliation{International Centre for Theory of Quantum Technologies, University of Gdansk, Gdansk, Poland}

\author{K. S. Mallesh} 
\affiliation{Regional Institute of Education (NCERT), Mysuru 570006, India}

\author{A. R. Usha Devi} 
\affiliation{Department of Physics, Bangalore University, 
	Bangalore-560 056, India}
\affiliation{Inspire Institute Inc., Alexandria, Virginia, 22303, USA.}

\begin{abstract} 
 The quantum steering ellipsoid inscribed inside the Bloch sphere offers an elegant geometric visualization of two-qubit states shared between  Alice and Bob. The set of  Bloch vectors of Bob's qubit, steered by Alice via  all possible local measurements on her qubit, constitutes the steering ellipsoid.   The steering ellipsoids are shown to be  effective in capturing quantum correlation properties, such as  monogamy,  exhibited by entangled multiqubit systems. We focus here on the canonical ellipsoids of two-qubit states realized by incorporating optimal local filtering operations by Alice and Bob on their respective qubits. Based on these canonical forms we show that the reduced two-qubit states drawn from  pure entangled three-qubit permutation symmetric states, which are inequivalent under  stochastic local operations and classcial communication (SLOCC), carry distinct geometric signatures. We provide detailed analysis of the SLOCC canonical forms and the associated steering ellipsoids of  the reduced two-qubit states extracted from entangled three-qubit pure symmetric states: We arrive at  (i) a prolate spheroid centered at the origin of the Bloch sphere -- with longest   semiaxis along the $z$-direction (symmetry axis of the spheroid) equal to  1 -- in the case of pure symmetric three-qubit states constructed by permutation of 3 distinct spinors and (ii) an oblate spheroid centered at $(0,0,1/2)$ inside  the Bloch sphere, with fixed semiaxes lengths $(1/\sqrt{2},\, 1/\sqrt{2},\, 1/2)$,  when the three-qubit pure state is constructed via symmetrization of 2 distinct spinors. We also explore volume monogamy relations formulated in terms of the volumes of the steering ellipsoids of the SLOCC inequivalent pure entangled three-qubit symmetric states.
\end{abstract}
\maketitle
\section{Introduction} 
 The Bloch sphere representation of a single qubit state provides a valuable geometric intuition in basic quantum information processing protocols. A natural generalization of the Bloch sphere picture to visualize two-qubit states, shared between Alice and Bob say, is given by the quantum steering ellipsoid~\cite{jevtic2014,MilneNJP2014,MilnePRA2016}. In particular, the set of all points  on the surface of the  ellipsoid  correspond to  Bloch vectors to which Bob's qubit can be steered to via all possible local measurements carried out on Alice's qubit. Suppose that Alice and Bob carry out local  local filtering operations on their respective qubits so as to reduce the two-qubit state into its SLOCC canonical forms: One of the SLOCC canonical forms  happens to be the Bell diagonal form of two-qubit state and the other, a nondiagonal canonical form~\cite{verstraete2001,supra}. The steering ellipsoids associated with the SLOCC canonical forms provide a much simpler geometric picture representing the set of all SLOCC equivalent two-qubit states. In this paper we explore  SLOCC canonical forms of the two-qubit reduced state, extracted from pure entangled three-qubit permutation symmetric state. We show that inequivalent SLOCC families of entangled three-qubit pure permutation symmetric states exhibit distinct canonical steering ellipsoids for the constituent two-qubit subsystem. In other words,  SLOCC classes of pure symmetric three-qubit entangled states can be classified based entirely on their canonical steering ellipsoids . We also show that the SLOCC canonical forms of the steering ellipsoids capture monogamy properties of the pure three-qubit entangled state effectively and they mirror insightful information about two-qubit entanglement.    

Contents of this paper are organized as follows:  In Sec.~2, we give an outline of the SLOCC canonical structure~\cite{supra} of an arbitrary two-qubit density matrix and  the associated canonical steering ellipsoid inscribed inside the Bloch sphere. Sec.3 is devoted to a brief review on SLOCC classification of pure permutation symmetric $n$-qubit states based on Majorana  representation~\cite{majorana,bastin,aru}. Explicit parametrization of two SLOCC inequivalent families of three-qubit entangled pure symmetric states~\cite{meyer} is given in Sec.~4.  We obtain the SLOCC canonical forms and the associated steering ellipsoid of the reduced  two-qubit states drawn from entangled three-qubit pure symmetric states in Sec.~5. Discussions on volume monogamy relation obeyed  by the three-qubit pure symmetric states belonging to  SLOCC inequivalent families and bounds imposed on the concurrence in terms of the obesity of the steering ellipsoid  is elucidated in Sec.~6. Summary of our results is presented in Sec.~7.  

 \section{ The quantum steering ellipsoid  of the two-qubit SLOCC canonical form} 
We  outline the procedure developed in Ref.~\cite{supra} to obtain  SLOCC canonical form  and the associated geometric visualization of an arbitrary two-qubit system.    

Consider a two-qubit density matrix $\rho_{AB}$, expanded in the Hilbert-Schmidt basis $\{\sigma_\mu\otimes \sigma_\nu, \mu,\nu=0,1,2,3\}$: 
\begin{eqnarray}
	\label{rho2q}
	\rho_{AB}&=&\frac{1}{4}\, \sum_{\mu,\,\nu=0}^{3}\,   
	\Lambda_{\mu \, \nu}\, \left( \sigma_\mu\otimes\sigma_\nu \right), \\ 
	\label{lambda}
	\Lambda_{\mu \, \nu}&=& {\rm Tr}\,\left[\rho_{AB}\,
	(\sigma_\mu\otimes\sigma_\nu)\,\right] =  	\Lambda^*_{\mu \, \nu} 
\end{eqnarray}
where
\begin{eqnarray}
	\label{sigmamu}
	\sigma_0=\left(\begin{array}{cc} 1 & 0 \\ 0 & 1       \end{array}\right),\ \  \sigma_1=\left(\begin{array}{cc} 0 & 1 \\ 1 & 0       \end{array}\right),\ \   \sigma_2=\left(\begin{array}{cc} 0 & -i \\ i & 0       \end{array}\right),\ \   \sigma_3=\left(\begin{array}{cc} 1 & 0 \\ 0 & -1       \end{array}\right).   
\end{eqnarray} 
The expansion coefficients $\Lambda_{\mu \, \nu}$ in (\ref{rho2q}) may be arranged as a $4\times 4$ real matrix 
\begin{eqnarray}
	\Lambda&=&\left(\begin{array}{llll} 1& r_1 & r_2 & r_3 \\ 
		s_1 & t_{11}  & t_{12} & t_{13} \\ 
		s_2 & t_{21}  & t_{22} & t_{23} \\
		s_3 & t_{31}  & t_{32} & t_{33} \\
	\end{array}\right).  
\end{eqnarray} 
Here ${\mathbf r}=(r_1,\,r_2,\,r_3)^T$, ${\mathbf s}=(s_1,\,s_2,\,s_3)^T$ are respective Bloch vectors of Alice, Bob's qubits and $T=(t_{ij})$ is the correlation matrix i.e., 
\begin{eqnarray} 
	\label{ri}
	r_i&=&\Lambda_{i \, 0}= {\rm Tr}\,\left[\rho_{AB}\, (\sigma_i\otimes\sigma_0)\,\right] \ \ \\
	\label{sj} 
	s_j&=& \Lambda_{0 \, j}={\rm Tr}\,\left[\rho_{AB}\, (\sigma_0\otimes\sigma_j)\,\right] \\ 
	\label{tij}
	t_{ij}&=& \Lambda_{i \, j}= {\rm Tr}\,\left[\rho_{AB}\, (\sigma_i\otimes\sigma_j)\,\right],\  \ \ \ i,\,j=1,\,2,\,3. 
\end{eqnarray} 
\begin{itemize}
	\item Under SLOCC transformation  the two-qubit density matrix  
	\begin{eqnarray}
		\rho_{AB}\longrightarrow\widetilde{\rho}_{AB}&=&\frac{(A\otimes B)\, \rho_{AB}\, (A^\dag\otimes B^\dag)}
		{{\rm Tr}\left[\rho_{AB}\, (A^\dag\, A\otimes B^\dag\, B)\right]}
	\end{eqnarray} 
	where $A, B\in {\rm SL(2,C)}$ denote $2\times 2$ complex matrices with unit determinant.
	\item Corresponding to the SLOCC transformation $\rho_{AB}\longrightarrow\widetilde{\rho}_{AB}$ of the two-qubit state, the $4\times 4$ real matrix $\Lambda$ transforms as 
	\begin{eqnarray}
		\label{sl2c}
		\Lambda\longrightarrow \widetilde{\Lambda}&=&\frac{L_A\,\Lambda\, L^T_B}{\left(L_A\,\Lambda\, L^T_B\right)_{00}}
	\end{eqnarray} 
	where $L_A,\, L_B\in SO(3,1)$ are $4\times 4$  proper orthochronous Lorentz transformation matrices~\cite{KNS} corresponding to  $A$, 
	$B\in SL(2,C)$ respectively and the superscript `$T$' denotes transpose operation.  
	\item  The $4\times 4$ real symmetric matrix $\Omega=\Lambda\, G\, \Lambda^T$, where $G={\rm diag}\,(1,-1,-1,-1)$ denotes the Lorentz metric, 
	undergoes a {\em Lorentz congruent transformation} under SLOCC (upto an overall  factor)~\cite{supra}:   
	\begin{eqnarray}
		\label{oa} 
		\Omega\rightarrow \widetilde{\Omega}_A&=& \widetilde{\Lambda}\, G\, \widetilde{\Lambda}^T \nonumber \\
		&=& L_{A}\, \Lambda\, L_{B}^T\, G \, L_{B}\, \Lambda^T L_{A}^T \nonumber \\ 
		&=& L_{A}\, \Omega\, L_{A}^T  
	\end{eqnarray}
	where the defining property~\cite{KNS}   $L^T\,G\,L=G$  of Lorentz transformation is used.  The real symmetric matrix  $\Omega$ plays a significant role in obtaining SLOCC canonical form and the corresponding geometrical structure of the two-qubit density matrix $\rho_{AB}.$ 
	\item It is seen that the matrix $G\,\Omega$, constructed using the real matrix parametrization  $\Lambda$ of a two-qubit density matrix $\rho_{AB}$ (see (\ref{rho2q}), (\ref{lambda})), exhibits the following properties~\cite{supra}:  
	\begin{itemize}
		\item[(i)] The $4\times 4$ matrix $G\,\Omega$  possesses  {\em non-negative} eigenvalues $\lambda_0\geq\lambda_1\geq\lambda_2\geq \lambda_3\geq 0$.  
		\item [(ii)]   Eigenvector $X$ associated with the highest eigenvalue $\lambda_0$ of  the matrix $G\,\Omega$ obeys one of the following  Lorentz invariant conditions: 
		\begin{equation}
			\label{positive}
			X^T\, G\, X>0   \\
		\end{equation}
		or
		\begin{equation}
			\label{neutral} 
			X^T\, G\, X =0.  
		\end{equation} 
		Furthermore the condition (\ref{neutral}) is accompanied by the observation that the matrix $G\Omega$ has only two eigenvalues $\lambda_0$, $\lambda_1$ obeying $ \lambda_0\geq \lambda_1$, both of which are  doubly degenerate.
		\item [(iii)] Suppose the eigenvector  $X$ satisfies the Lorentz invariant condition (\ref{positive}). Then there exists  suitalbe SLOCC transformations $A_1,\, B_1\in SL(2,C)$ (with corresponding    Lorentz transformations $L_{A_1},\, L_{B_1}\in SO(3,1)$ respectively) such that the matrix $\Omega$ assumes a diagonal form: 
		\begin{eqnarray} 
			\label{o1c}
			\widetilde{\Omega}_{I_c}&=& L_{A_{I_c}}\, \Omega\, L^T_{A_{I_c}}  ={\rm diag}  \,  \left(\lambda_0,\, -\lambda_1,\, -\lambda_2,\, -\lambda_3\right).   
		\end{eqnarray}
		Correspondingly, one obtains the Lorentz canonical form for the real matrix $\Lambda$:       
		\begin{eqnarray} 
			\label{lambda1c}
			\Lambda\longrightarrow\widetilde{\Lambda}_{I_c}&=&\frac{L_{A_1}\,\Lambda\, L^T_{B_1}}{\left(L_{A_1}\,\Lambda\, L^T_{B_1}\right)_{00}} =  {\rm diag}\,  \left(1,\,\sqrt{\frac{\lambda_1}{\lambda_0}},\sqrt{\frac{\lambda_2}{\lambda_0}},\, \pm\, \sqrt{\frac{\lambda_3}{\lambda_0}}\right) 
		\end{eqnarray}
		where the sign $\pm$ is chosen depending on   ${\rm sign}\left( \det(\Lambda)\right)=\pm$ and 
		\begin{equation}
			\label{l0}
		\left(L_{A_1}\,\Lambda\, L^T_{B_1}\right)_{00}=\sqrt{\lambda_0}.
		\end{equation} 
		Consequently, the two-qubit density matrix 
		\begin{eqnarray*}	 
			\rho_{AB}\longrightarrow\widetilde{\rho}^{\,I_c}_{AB}&=&\frac{(A_{I_c}\otimes B_{I_c})\, \rho_{AB}\, (A_{I_c}^\dag\otimes B_{I_c}^\dag)}
			{{\rm Tr}\left[\rho_{AB}\, (A_{I_c}^\dag\, A_{I_c} \otimes B_{I_c}^\dag\, B_{I_c})\right]}
		\end{eqnarray*} 
		reduces to the Bell-diagonal form  
		\begin{eqnarray}	 
			\label{rhobd} 
			\widetilde{\rho}^{\,I_c}_{AB}
			&=&  \frac{1}{4}\, \left( \sigma_0\otimes \sigma_0 + \sum_{i=1,2}\, \sqrt{\frac{\lambda_i}{\lambda_0}}\, \sigma_i\otimes\sigma_i \pm \sqrt{\frac{\lambda_3}{\lambda_0}}\, \sigma_3\otimes\sigma_3 \right)
		\end{eqnarray}
		under SLOCC transformations. 
		\item [(iv)] Whenever the eigenvector  $X$ obeys the condition (\ref{neutral})  suitable SLOCC transformations 
		$A_{II_c},\, B_{II_c}\in SL(2,C)$ (associated  Lorentz transformations denoted respectively by $L_{A_{II_c}},\, L_{B_{II_c}}\in SO(3,1)$) exist such that    
		\begin{eqnarray}
			\label{o2c}
			\widetilde{\Omega}_{II_c}&=& L_{A_{II_c}}\, \Omega\, L^T_{A_{II_c}} =      
			\left(\begin{array}{cccc}
				\phi_0 & 0  & 0 & \phi_0-\lambda_0 \\ 
				0 & -\lambda_1 & 0 & 0 \\ 
				0 & 0 &   -\lambda_1 & 0 \\
				\phi_0-\lambda_0 & 0 & 0 &  \phi_0-2\,\lambda_0, 
			\end{array}\right),         
		\end{eqnarray}
		where 
		\begin{equation}
			\phi_0=\left(L_{A_{II_c}}\,\Omega\,L_{A_{II_c}}^T\right)_{00}.
		\end{equation}
		As a consequence, the real matrix $\Lambda$ and the two-qubit density matrix $\rho_{AB}$ assume  the following  canonical forms:        
		\begin{eqnarray} 
			\label{lambda2c}
			\Lambda\longrightarrow\widetilde{\Lambda}_{II_c}&=&\frac{L_{A_{II_c}}\,\Lambda\, L^T_{B_{II_c}}}{\left(L_{A_{II_c}}\,\Lambda\, L^T_{B_{II_c}}\right)_{00}}    
			= \left(\begin{array}{cccc}
				1 & 0  & 0 & 0 \\ 
				0 & a_1 & 0 & 0 \\ 
				0 & 0 &   -a_1 & 0 \\
				1-a_0 & 0 & 0 &  a_0 
			\end{array}\right)
		\end{eqnarray}
		where 
		\begin{equation}
			\label{phi0}
			\left(L_{A_{II_c}}\,\Lambda\, L^T_{B_{II_c}}\right)_{00}=\sqrt{\phi_0}.
		\end{equation}
		\begin{eqnarray}
			\label{rho2c} 
			\rho_{AB}\longrightarrow\widetilde{\rho}^{\,II_c}_{AB}&=&\frac{(A_{II_c}\otimes B_{II_c})\, \rho_{AB}\, (A_{II_c}^\dag\otimes B_{II_c}^\dag)}
			{{\rm Tr}\left[\rho_{AB}\, (A_{II_c}^\dag\, A_{II_c} \otimes B_{II_c}^\dag\, B_{II_c})\right]},  \\ 
			&=&  \frac{1}{4}\, \left[ \, \sigma_0\otimes \sigma_0 + (1-a_0)\,\sigma_3\otimes \sigma_0 + a_1 \,(\sigma_1\otimes\sigma_1 - \sigma_2\otimes\sigma_2) +\,a_0\, \sigma_3\otimes\sigma_3\right]  \nonumber  
		\end{eqnarray}
		where 
		\begin{equation}
			\label{adef}
			a_0=\frac{\lambda_0}{\phi_0},\ \ \ \ \  a_1=\sqrt{\frac{\lambda_1}{\phi_0}}, \ \  \ 0\leq a_1^2\leq a_0\leq 1. 
		\end{equation}
 \item[(v)]  Denote the set of all  projective valued measurements (PVM) on Bob's qubit  by    $$\left\{E=\sum_{\mu=0}^{3}\, e_\mu\, \sigma_\mu, E>0,\ e_0=1, {\mathbf e}=(e_1,e_2,e_3),\ \ {\mathbf e}\cdot{\mathbf e}=e_1^2+e_2^2+e_3^2=1\right\}.$$ Local PVM on the Bob's qubit  leads to collapsed state of Alice's qubit 
		$$\rho_{\mathbf p}=\frac{1}{p_E}{\rm Tr}_B\,\left[\rho_{AB}\,\left(\sigma_0\otimes E\right)\right]=\frac{1}{2}\sum_{\mu}\, p_\mu\, \sigma_\mu,\ \ p_\mu= \frac{1}{p_E}\,\sum_\nu\,\Lambda_{\mu\,\nu}\,e_\nu,$$ 
	with probability
		\begin{eqnarray*}
		p_{E}&=&{\rm Tr}\,\left[\rho_{AB}\,\left(\sigma_0\otimes E\right)\right]=\sum_{\mu=0}^{3}\,\Lambda_{0\,\mu}\,e_\mu= 1+\mathbf{r}\cdot\mathbf{e}.
	\end{eqnarray*} 
Steered  Bloch vectors of Alice's qubit lie on and within the Bloch sphere and they constitute an ellipsoidal surface defined by 
 ${\mathcal E}_{A\vert\,B}=\left\{\mathbf{p}=\frac{\mathbf{s}\,+\,T^{T}\,\mathbf{e}}{1+\,\mathbf{r}\cdot\,\mathbf{e}}\right\}.$ Points inside the  steering ellipsoid are accessible when Bob employs convex combinations of PVMs i.e., positive operator values measures (POVMs) to steer  Alice's qubit.  

\item[(vi)]For the Lorentz canonical form $\widetilde{\Lambda}_{I_c}$  (see (\ref{lambda1c})) of the two-qubit state $\widetilde{\rho}^{\,I_c}_{AB}$ one obtains  
		\begin{equation}
		\mathbf{p}=\left(p_1=\sqrt{\frac{\lambda_1}{\lambda_0}}\, e_1,\, p_2=\sqrt{\frac{\lambda_2}{\lambda_0}}\, e_2,\, \sqrt{\frac{\lambda_3}{\lambda_0}}\, e_3\right)
			\end{equation} 
	as steered Bloch points of Alice's qubit, which obey  the equation 
		$$\frac{\lambda_0\, p_1^2}{\lambda_1}+ \frac{\lambda_0\, p_2^2}{\lambda_1}+ \frac{\lambda_0\, p_3^2}{\lambda_1}=1$$
		of an  ellipsoid with semi-major axes  $(\sqrt{\lambda_1/\lambda_0}, \,\sqrt{\lambda_2/\lambda_0}\, \sqrt{\lambda_3/\lambda_0})$ and center  $(0,0,0)$ inside the Bloch sphere. We refer to this as the {\em canonical steering ellipsoid} of the set of all two-qubit density matrices which are on the SLOCC orbit of the Bell-diagonal state $\widetilde{\rho}^{\,I_c}_{AB}$ (see (\ref{rhobd})). 
		
		\item[(vii)] Corresponding to the second Lorentz canonical form   $\widetilde{\Lambda}_{II_c}$ (see (\ref{lambda2c})) one obtains canonical steering spheroid inside the Bloch sphere with its semiaxes lengths $(a_1,a_1, a_0)$  and center $(0,\,0,\, 1-a_0)$.    In other words a shifted spheroid, inscribed within the Bloch sphere, represents  two-qubit states, which are SLOCC equivalent to $\widetilde{\rho}_{AB}^{II_c}$ (see  (\ref{rho2c})). One obtains a prolate spheroid,  centered at the origin of the Bloch sphere, with semiaxes length equal to 1 along $z$-axis when $a_0=1$ as the canonical steering ellipsoid offering geometric visualization of the  set of all two-qubit states on the SLOCC orbit of $\widetilde{\rho}_{AB}^{II_c}$ given by (\ref{rho2c})).     
\end{itemize} 
\end{itemize}

\section{SLOCC classification of pure permutation symmetric states}

In 1932 Majorana~\cite{majorana}  expressed an arbitrary pure  symmetric state $\vert \Psi_{\rm sym}\rangle$  of spin $j=\frac{N}{2}$
 as a  {\em symmetrized} combination  of  $N$ constituent spinors (qubits) as follows: 
\begin{equation}
	\label{Maj}
	\vert \Psi_{\rm sym}\rangle={\cal N}\, \sum_{P}\, \hat{P}\, \{ \vert \alpha_1, \beta_1\, \rangle \otimes  |\alpha_2, \beta_2\rangle\otimes\ldots \otimes \vert \alpha_N, \beta_N\rangle \}
\end{equation} 
where
$	 \vert \alpha_k, \beta_k\, \rangle =\cos\frac{\beta_k}{2}\,\vert 0\rangle
	+e^{i\alpha_k}\,\sin\frac{\beta_k}{2}\,  \vert 1\rangle,$\ $0\leq \beta_k\leq \pi,$  $0\leq \alpha_k\leq 2\pi$
 denote the constituent spinor states (states of the  qubits). 
Here $\hat{P}$ denotes the set of all $N!$ 
permutations and ${\cal N}$ corresponds to an overall normalization factor.  The parameters  $(\alpha_k,\beta_k)$
characterizing the states $\vert \alpha_k, \beta_k\, \rangle , \ \  k=1,2,\ldots, N$ of the  constituent qubits  offer  geometric representation of $N$-qubit pure symmetric state $\vert \Psi_{\rm sym}\rangle$ in terms of  $N$ points on the Bloch sphere.  This is referred to as the Majorana representation of an arbitrary $N$-qubit symmetric 
state $\vert \Psi_{\rm sym}\rangle$  in terms of the constituent qubit states $ \vert \alpha_k, \beta_k\, \rangle ,\ k=1,2,\ldots, N$.

Recall that two $N$-qubit   states $\vert\Phi\rangle$, $\vert \chi\rangle$  can be obtained from one another by means of stochastic local operations and classical communications (SLOCC)  if and only if there exists a local operation  $A_1\otimes A_2\otimes \ldots \otimes A_N$, where $A_k, k=1,2,\ldots, N$ denote  $2\times 2$ complex invertible matrices, such that $\vert \Phi\rangle=A_1\otimes A_2\otimes \ldots \otimes A_N\vert\chi\rangle.$   In the special case of permutation symmetric $N$-qubits it is sufficient 
to search for identical local operations of the form $A^{\otimes N}=A\otimes A\otimes\ldots \otimes A$ to verify the SLOCC equivalence~\cite{bastin,aru}.   The Majorana representation discussed above leads to a  natural identification of  different SLOCC inequivalent families, depending on the number and arrangement of  the distinct qubits constituting the pure symmetric state~\cite{bastin,aru}. From early studies it was identified that the three-qubit GHZ and W states are inequivalent under SLOCC~\cite{dur}. Based on Majorana geometric representation it is realized that GHZ class of 
three-qubit pure symmetric states are constituted by 3 distinct qubits whereas W class of states consist of 2 distinct qubits. In Ref.~\cite{meyer} Meill and Meyer proposed a convenient parametrization,  based on the Majorana geometric representation~\cite{bastin,aru}, and evaluated the algebraically independent local unitary invariant quantities  of these SLOCC inequivalent classes of  pure entangled three-qubit symmetric   states. More recently, some of us~\cite{anjali} carried out a detailed analysis on the non-locality features of entangled pure symmetric three-qubit states, belonging to SLOCC inequivalent classes  using the Meill-Meyer parametrization~\cite{meyer}. In the following sections we study the explicit structure of steering ellipsoids of SLOCC canonical forms~\cite{verstraete2001,supra} of the reduced two-qubit density matrices, extracted from entangled pure symmetric three-qubit states. We show that the two SLOCC inequivalent families of pure entangled three-qubit symmetric states exhibit distinct steering ellipsoids for their reduced two-qubit density matrices. 
 
\section{SLOCC classification of pure symmetric three-qubit  states} 
A pure symmetric three-qubit  state can be expressed in the Majorana geometric representation (see (\ref{Maj})) as     
\begin{eqnarray}
\label{psisym1}
\vert\Psi^{ABC}_{\rm sym}\rangle&=&{\cal N} \sum_P \hat{P}\{ \vert \alpha_1, \beta_1\, \rangle \otimes  |\alpha_2, \beta_2\rangle\otimes \vert \alpha_3, \beta_3\rangle \}
\end{eqnarray} 
where, 
$\vert \alpha_k, \beta_k\rangle\equiv \cos\frac{\beta_k}{2}\,  
\vert 0\rangle + e^{i\alpha_k}\,\sin\frac{\beta_k}{2}\, \vert 1\rangle,\ \ k=1,\,2,\,3.$
are the states of the constituent qubits.

Depending on the number of distinct spinor states $\{ \vert \alpha_k, \beta_k\,\rangle,\, k=1,2,3\}$ there arise {\em two} different classes of {\em entangled} pure symmetric states of three-qubits~\cite{bastin,aru,anjali}:    
\begin{enumerate}
	\item Two distinct spinor class:  $\vert\alpha_1,\beta_1\rangle =\vert\alpha_2,\beta_2\rangle\neq \vert\alpha_3,\beta_3\rangle$ 
	\item Three distinct spinor class:  $\vert\alpha_1,\beta_1\rangle \neq \vert\alpha_2,\beta_2\rangle \neq   
	\vert\alpha_3,\beta_3\rangle$. 
\end{enumerate} 
Note that if there is only one distinct spinor characterizing the state i.e., when $\vert\alpha_1,\beta_1\rangle =\vert\alpha_2,\beta_2\rangle= \vert\alpha_3,\beta_3\rangle=\vert\alpha,\beta\rangle$, the pure  state $\vert\Psi^{ABC}_{\rm sym}\rangle\equiv \vert \Psi^{ABC}_{3,1}\rangle$ of (\ref{psisym1})  reduces to  
\be
\label{sepproduct}
\vert \Psi^{ABC}_{3,1}\rangle= \vert\,\alpha,\beta\rangle \otimes \vert\,\alpha,\beta\rangle \otimes \vert\,\alpha,\beta\rangle
=\vert\,\alpha,\beta\rangle^{\otimes\,3} 
\ee
and hence it is separable. 

We denote the two distinct spinor class of pure three-qubit symmetric states by ${\cal D}_{3,2}$, the three distinct spinor class by ${\cal D}_{3,3}\}$ and the one distinct spinor class (unentangled) by 
${\cal D}_{3,1}$.  These three classes are evidently inequivalent under SLOCC~\cite{bastin,aru}. Three-qubit W state given by   
\begin{eqnarray}
\label{W}
\vert{\rm W}\rangle&=& {\cal N}\, \sum_P \hat{P}\{ \vert 0\,\rangle \otimes  |0\,\rangle\otimes \vert 1\,\rangle \}\nonumber \\
&=& \frac{1}{\sqrt{3}}\, \left(\vert 0_A\,0_B\, 1_C\rangle+\vert 0_A\,1_B\, 0_C\rangle+\vert 1_A\,0_B\, 0_C\rangle\right)
\end{eqnarray}
and  its obverse state 
\begin{eqnarray}
	\label{Wbar}
	\vert\overline{{\rm W}}\rangle&=& {\cal N}\, \sum_P \hat{P}\{ \vert 1\,\rangle \otimes  |1\,\rangle\otimes \vert 0\,\rangle \}\nonumber \\
	&=& \frac{1}{\sqrt{3}}\, \left(\vert 1_A\,1_B\, 0_C\rangle+\vert 1_A\,0_B\, 1_C\rangle+\vert 0_A\,1_B\, 1_C\rangle\right)
\end{eqnarray}
belong to the class  ${\cal D}_{3,2}$.  
Three-qubit GHZ state is a well-known example of permutation symmetric state belonging to the SLOCC class  ${\cal D}_{3,3}$:  
\begin{eqnarray}
\label{GHZ}
\vert{\rm GHZ}\rangle&=& {\cal N}\, \sum_P \hat{P}\{ \vert\, \phi\rangle_1 \,  \otimes \vert\, \phi\rangle_2  \otimes \vert \phi\rangle_3 \}\nonumber \\
&=& \frac{1}{\sqrt{2}}\, \left(\vert 0_A\,0_B\, 0_C\rangle+\vert 1_A\,1_B\, 1_C\rangle\right) 
\end{eqnarray}
where    $\vert\phi\rangle_p=\frac{1}{\sqrt{2}}(\vert 0\rangle+\omega^p\,\vert 1\rangle), p=1,2,3$ are the constituent qubit states; $\omega$ denotes the cube root of unity. 
 
An arbitrary three-qubit pure state $\vert\Psi^{ABC}_{3,\,2}\rangle$ belonging to the two distinct spinor class ${\cal D}_{3,2}$ takes the following simple form  under local unitary operations~\cite{aru,meyer}: 
\begin{eqnarray}
\label{2meyer}
\vert\Psi^{ABC}_{3,\,2}\rangle&=&{\cal N}_{3,\,2} \sum_P \hat{P}\{ \vert 0\rangle \otimes  \vert 0\rangle \otimes \vert \beta\rangle\},  \hskip 0.2in \vert \beta\rangle=\cos\,\frac{\beta}{2}\, \vert 0\rangle+\sin\,\frac{\beta}{2}\,\vert 1\rangle, \ 0<\beta\leq \pi \nonumber \\ 
&=& \frac{1}{\sqrt{2+\cos\beta}}\, \left( \sqrt{3}\,\cos\frac{\beta}{2}\,\left\vert 0_A\,0_B\, 0_C\,\right\rangle + \sin\frac{\beta}{2}\,\left\vert{\rm W}\,\right\rangle\right).
\end{eqnarray}	
In other words the states belonging to the two distinct spinor class ${\cal D}_{3,2}$ are characterized by {\em one} real parameter $\beta$. 

The reduced two-qubit density matrix extracted from the pure state $\vert\Psi^{ABC}_{3,\,2}\rangle$ is expressed (in the computational basis  $\{\vert 0, 0\rangle, \vert 0\, 1\rangle, \vert 1\, 0\rangle, \vert 1\, 1\rangle \}$) by
\begin{eqnarray}
	\label{rrho32}
	\rho^{(2-{\rm qubit})}_{3,\,2} &=& {\rm Tr}_{A}\,\vert\Psi^{ABC}_{3,\,2}\rangle\langle \Psi^{ABC}_{3,\,2}\vert = {\rm Tr}_{B}\,\vert\Psi^{ABC}_{3,\,2}\rangle\langle \Psi^{ABC}_{3,\,2}\vert
	={\rm Tr}_{C}\,\vert\Psi^{ABC}_{3,\,2}\rangle\langle \Psi^{ABC}_{3,\,2}\vert \nonumber \\
&=& \ba{cccc} 1-2A_{3,2}  & B_{3,2} &   B_{3,2} & 0 \\ 
              B_{3,2} & A_{3,2} & A_{3,2} & 0  \\ 
              B_{3,2} & A_{3,2} & A_{3,2} & 0 \\ 
              0 & 0 & 0 & 0        \ea   
\end{eqnarray}
where 
\begin{equation}
	\label{rrho32e}
	A_{3,2}=\frac{1-\cos\beta}{6\,(2+\cos \beta)}, \ \ B_{3,2}=\frac{\sin\beta}{2\,(2+\cos \beta)}. 
\end{equation}

Similarly, any arbitrary three-qubit pure symmetric state $\vert\Psi^{\rm ABC}_{3,3}\rangle$ belonging to Majorana three distinct spinor class  ${\cal D}_{3,3}\}$  can be reduced -- under local unitary operations --  to the simple form given by~\cite{meyer,anjali}
\begin{eqnarray}
\label{3meyer} 
\vert\Psi^{\rm ABC}_{3,3}\rangle&=&{\cal N}_{3,\,3} \left(\vert 0\,\rangle^{\otimes\,3}  + y\,  
e^{i\, \alpha} \, \vert \beta\rangle^{\otimes\,3} \right),\hskip 0.1in \vert \beta\rangle=\cos\,\frac{\beta}{2}\, \vert 0\rangle+\sin\,\frac{\beta}{2}\,\vert 1\rangle, 
\ \ 0<y\leq 1, \ 0\leq \alpha \leq 2\pi,\ 0<\beta\leq \pi, \nonumber \\
&=& {\cal N}_{3,\,3} \left[\left(1+y\,e^{i\alpha}\,\cos\frac{\beta}{2}\right)\, \vert 0_A\,0_B\, 0_C\rangle +  y\,e^{i\alpha}\,\sin\frac{\beta}{2} \vert 1_A\,1_B\, 1_C\rangle
+ \frac{\sqrt{3}}{2}\,y\,\sin\beta\,\left(\cos\frac{\beta}{2}\,\vert{\rm W}\rangle+ \sin\frac{\beta}{2}\,\vert\overline{{\rm W}}\rangle\right) \right]  \nonumber \\  
\end{eqnarray} 
and is characterized by three real parameters $y, \alpha$ and $\beta$.  
 
 The two-qubit density matrix drawn from the pure state $\vert\Psi^{ABC}_{3,\,3}\rangle$ (see (\ref{3meyer})) is given explicitly by (in the computational basis  $\{\vert 0, 0\rangle, \vert 0\, 1\rangle, \vert 1\, 0\rangle, \vert 1\, 1\rangle \}$) 
 \begin{eqnarray}
 	\label{rrho33}
 	\rho^{(2-{\rm qubit})}_{3,\,3} &=& {\rm Tr}_{A}\,\vert\Psi^{ABC}_{3,\,3}\rangle\langle \Psi^{ABC}_{3,\,3}\vert = {\rm Tr}_{B}\,\vert\Psi^{ABC}_{3,\,3}\rangle\langle \Psi^{ABC}_{3,\,3}\vert
 	={\rm Tr}_{C}\,\vert\Psi^{ABC}_{3,\,3}\rangle\langle \Psi^{ABC}_{3,\,3}\vert \nonumber \\ 
	& & \nonumber \\ 
	& & \nonumber \\
 	&=& \ba{cccc} A_{3,3}  & B_{3,3} &   B_{3,3} & C_{3,3} \\ 
 	B^*_{3,3} & D_{3,3} & D_{3,3} &   E_{3,3}  \\ 
 	B^*_{3,3} & D_{3,3} & D_{3,3} & E_{3,3}\\ 
 	 C^*_{3,3} & E^*_{3,3} & E^*_{3,3} & F_{3,3}       \ea   
 \end{eqnarray}
 where 
 \begin{eqnarray}
 	\label{rrho33e}
 	A_{3,3}&=& \frac{1+y^2\,\cos^2\frac{\beta}{2}+2\,y\, \cos\alpha\,\cos^3\frac{\beta}{2}}{1+y^{2}+2y\,\cos\alpha\,\cos^3\frac{\beta}{2}}, \ \ \ B_{3,3}=\frac{ y\,\cos^2\frac{\beta}{2}\, \sin\frac{\beta}{2}\,\left( e^{- i\alpha}+y\,\cos\frac{\beta}{2}\right)}{1+y^{2}+2y\,\cos\alpha\,\cos^3\frac{\beta}{2}},\nonumber \\ 
 	 C_{3,3}&=&\frac{ y\,\sin^2\frac{\beta}{2}\, \cos\frac{\beta}{2}\,\left( e^{- i\alpha}+y\,\cos\frac{\beta}{2}\right)}{1+y^{2}+2y\,\cos\alpha\,\cos^3\frac{\beta}{2}}, \ \ \ 
 	D_{3,3}=\frac{ y^2\,\sin^2\frac{\beta}{2}\, \cos^2\frac{\beta}{2}}{1+y^{2}+2y\,\cos\alpha\,\cos^3\frac{\beta}{2}}, \\ 
 	 E_{3,3}&=& \frac{ y^2\,\sin^3\frac{\beta}{2}\, \cos\frac{\beta}{2}}{1+y^{2}+2y\,\cos\alpha\,\cos^3\frac{\beta}{2}}, \hskip 0.5in  \ \ F_{3,3}=1-A_{3,3}-2\,D_{3,3}. \nonumber
 \end{eqnarray}
 We make use of the explicit form of the two-qubit density matrices $\rho^{(2-{\rm qubit})}_{3,\,2}$ (see (\ref{rrho32}), (\ref{rrho32e})) and $\rho^{(2-{\rm qubit})}_{3,\,3}$ (see (\ref{rrho33}),\,(\ref{rrho33e})) to compute the SLOCC canonical forms and the associated geometrical representations  following the detailed analysis given in  Ref.~\cite{supra}.

 \section{Canonical steering ellipsoids of $\rho^{(2-{\rm qubit})}_{3,\,2}$ and $\rho^{(2-{\rm qubit})}_{3,\,3}$}   
	 
In this section we focus our attention on determining the SLOCC canonical forms and the associated quantum steering ellipsoids of the two-qubit density matrices $\rho^{(2-{\rm qubit})}_{3,\,2}$ (see  (\ref{rrho32}), (\ref{rrho32e})) and $\rho^{(2-{\rm qubit})}_{3,\,3}$  (see \ref{rrho33}), (\ref{rrho33e}))  which are obtained by tracing out one of the qubits from the respective three-qubit pure states 	$\vert\Psi^{ABC}_{3,\,2}\rangle$, $\vert\Psi^{ABC}_{3,\,3}\rangle$ given by (\ref{2meyer}), (\ref{3meyer}).

\subsection{SLOCC Canonical form of $\bf{\rho^{(2-{\rm qubit})}_{3,\,2}}$}
Expressing the density matrix $\rho^{(2-{\rm qubit})}_{3,\,2}$ given by (\ref{rrho32}), (\ref{rrho32e}) in  the Hilbert-Schmidt basis $\{\sigma_\mu\otimes \sigma_\nu, \mu,\nu=0,1,2,3\}$ we obtain  
\begin{eqnarray*}
	\rho^{(2-{\rm qubit})}_{3,\,2}&=&\frac{1}{4}\, \sum_{\mu,\,\nu=0}^{3}\,   
	\Lambda^{(3,\,2)}_{\mu \, \nu}\, \left( \sigma_\mu\otimes\sigma_\nu \right), \\
	\Lambda^{(3,\,2)}_{\mu \, \nu}&=&{\rm Tr}\,\left[\rho^{(2-{\rm qubit})}_{3,\,2}\,
	(\sigma_\mu\otimes\sigma_\nu)\,\right],  
\end{eqnarray*}
where the $4\times 4$ real matrix $\Lambda^{(3,\,2)}$ has the explicit structure 
\begin{eqnarray}
		\label{lambda32}
\Lambda^{(3,\,2)}&=&\left(\begin{array}{cccc}1 & \frac{\sin\beta}{2+\cos \beta} &0& \frac{5+4\cos\beta}{3(2+\cos\beta)}  \\ 
\frac{\sin\beta}{2+\cos \beta}& \frac{1-\cos\beta}{3(2+\cos\beta)} & 0  &  \frac{\sin\beta}{2+\cos \beta} \\ 
0 & 0 &  \frac{1-\cos\beta}{3(2+\cos\beta)}  & 0 \\ 
\frac{5+4\cos\beta}{3(2+\cos\beta)}  &  \frac{\sin\beta}{2+\cos \beta} & 0 &  \frac{4+5\cos\beta}{3(2+\cos\beta)}\end{array} \right)= \left(\Lambda^{(3,\,2)}\right)^T. 
\end{eqnarray} 
Following the procedure outlined in Sec.~2, we evaluate the $4\times 4$ symmetric matrix $\Omega^{(3,\,2)}$ from $\Lambda^{(3,\,2)}$: 
\begin{eqnarray}
\label{lambda32}
\Omega^{(3,\,2)}&=&\Lambda^{(3,\,2)}\, G\, \left(\Lambda^{(3,\,2)}\right)^T= \Lambda^{(3,\,2)}\, G\, \Lambda^{(3,\,2)}\nonumber \\ 
&=& \left(\begin{array}{cccc} 
	2\,u(\beta) & 0 &0 & u(\beta) \\
	0& -u(\beta) & 0  &  0 \\ 
	0 & 0 &  -u(\beta) & 0 \\ 
    u(\beta)  &  0 & 0 & 0\end{array} \right),\ \ \ \ u(\beta)=\left[\frac{1-\cos\beta}{3(2+\cos\beta)}\right]^2. 
\end{eqnarray} 
The matrix $\Omega^{(3,\,2)}$ is already in the canonical form given by (\ref{o2c}) with 
\begin{eqnarray}
	\label{lphi}
	\lambda_0=u(\beta)=\lambda_1,\  \ {\rm and}\ \  \phi_0=2\, u(\beta). 
\end{eqnarray}
Indeed  the eigenvalues of the matrix $G\,\Omega^{(3,\,2)},\ G={\rm diag}\, (1,\,-1,\,-1,\,-1)$ are four-fold degenerate: 
\begin{equation}
	\label{ev32}
	\lambda_0= u(\beta)=\left[\frac{1-\cos\beta}{3(2+\cos\beta)}\right]^2= \lambda_1=\lambda_2=\lambda_3
\end{equation}
 and one of the eigenvectors $X^{(3,2)}=(1,\, 0,\, 0,\, -1)^T$  obeys the Lorentz invariant condition (\ref{neutral}) confirming our observation. Substituting (\ref{lphi}) in (\ref{lambda2c}),\  (\ref{adef}) we find that  the  Lorentz canonical form  
 $\widetilde{\Lambda}^{(3,\,2)}$  of the real matrix  $\Lambda^{(3,\,2)}$ is independent of the state parameter $\beta$:  
 \begin{eqnarray}
 	\label{l32c}
 	\widetilde{\Lambda}^{(3,\,2)}&=&\left(\begin{array}{cccc}1 & 0 &0& 0  \\ 
 		0& \frac{1}{\sqrt{2}} & 0  &  0\\ 
 		0 & 0 &  -\frac{1}{\sqrt{2}}  & 0 \\ 
 		\frac{1}{2}  &  0 & 0 &  \frac{1}{2}\end{array} \right). 
 \end{eqnarray} 
Thus, we conclude that the reduced two-qubit density matrices extracted from the pure symmetric three-qubit state $\vert\Psi^{ABC}_{3,\,2}\rangle$ given by (\ref{2meyer}) can be reduced to the following SLOCC canonical form (see (\ref{rho2c}))
  \begin{eqnarray}
 	\label{rr32c}
 	\widetilde{\rho}^{\,(2-{\rm qubit})}_{3,\,2}&=& \frac{1}{4}\, \left[ \, \sigma_0\otimes \sigma_0 + \frac{1}{2}\,\sigma_3\otimes \sigma_0 + \frac{1}{\sqrt{2}} \,(\sigma_1\otimes\sigma_1 - \sigma_2\otimes\sigma_2) +\,\frac{1}{2}\, \sigma_3\otimes\sigma_3\right]. 
 \end{eqnarray} 
The quantum steering ellipsoid associated with the two-qubit state (\ref{rr32c}) is a spheroid centered at $(0,0,1/2)$ inside  the Bloch sphere, with fixed semiaxes lengths $(1/\sqrt{2},\, 1/\sqrt{2},\, 1/2)$  (See Fig.~1). It is interesting to note that the SLOCC canonical form is independent of the state parameter $\beta$. 
\begin{figure}[h]
	\begin{center}
		\includegraphics*[width=3.5in,keepaspectratio]{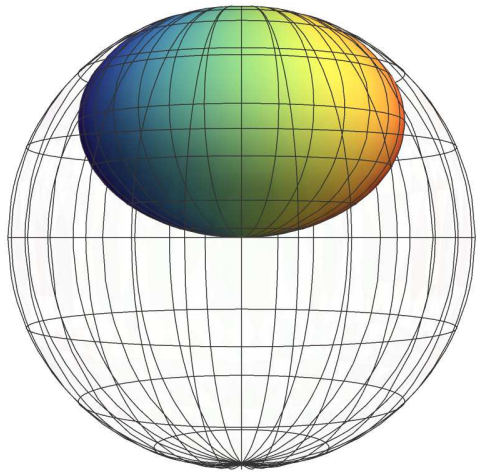}
		\caption{(Colour online) Steering spheroid inscribed within the Bloch sphere representing the Lorentz canonical form 	$\widetilde{\Lambda}^{(3,\,2)}$ (see (\ref{l32c})) geometrically. Lengths of the  semi-major axes of the shifted spheroid, which is centered at $(0,0,1/2)$, are given by $(\frac{1}{\sqrt{2}},\,\frac{1}{\sqrt{2}},\,\frac{1}{2})$}
	\end{center}
\end{figure}

  \subsection{SLOCC Canonical form of $\bf{\rho^{(2-{\rm qubit})}_{3,\,3}}$}
 From the explicit form of the density matrix  $\rho^{(2-{\rm qubit})}_{3,\,3}$ given by (\ref{rrho33}), (\ref{rrho33e}) we evaluate the elements \break $\Lambda^{(3,\,3)}_{\mu\,\nu}={\rm Tr}\,\left[\rho^{(2-{\rm qubit})}_{3\,3}\,
 (\sigma_\mu\otimes\sigma_\nu)\,\right],\ \mu,\nu=0,1,2,3$ of the  $4\times 4$ real matrix $\Lambda^{(3,\,3)}$:   
\begin{eqnarray}
	\label{l33e} 
\Lambda^{(3,\,3)}_{00} &=& 1, \ \Lambda^{(3,\,3)}_{01}= {\cal A}(y,\alpha,\beta)\,  \, y\,\sin\beta\,\left(y+\cos \alpha\, \cos \frac{\beta}{2}\right)  =\Lambda_{10}^{(3,\,3)}, \nonumber \\
	\Lambda_{02}^{(3,\,3)}&=& {\cal A}(y,\alpha,\beta)\, y\, \, \sin \alpha\,\sin\beta\, \cos\frac{\beta}{2}=\Lambda_{20}^{(3,\,3)}, \ \ 
	\Lambda_{03}^{(3,\,3)}= {\cal A}(y,\alpha,\beta)\, \left(1+2\,y\, \cos \alpha\, \cos^3\frac{\beta}{2}+y^2\,\cos \beta \right)=\Lambda_{30}^{(3,\,3)},   \nonumber \\ 
	\Lambda_{11}^{(3,\,3)}&=& {\cal A}(y,\alpha,\beta)\, y\,\sin\beta\, \sin \frac{\beta}{2}\,\left(\cos \alpha+2\,y\, \cos \frac{\beta}{2}\right), \ \    
	\Lambda_{12}^{(3,\,3)}={\cal A}(y,\alpha,\beta)\,y \sin \alpha\,\sin\beta\, \sin\frac{\beta}{2}=\Lambda_{21}^{(3,\,3)}, \nonumber \\
	\Lambda_{13}^{(3,\,3)}&=&{\cal A}(y,\alpha,\beta)\, y\,\sin \beta\, \left(\cos \alpha\, \cos \frac{\beta}{2}+y\, \cos \beta\right)=\Lambda_{31}^{(3,\,3)}, \ \ 
	 \Lambda_{22}^{(3,\,3)}=-{\cal A}(y,\alpha,\beta)\, y\,\cos\alpha\, \sin \beta\, \sin \frac{\beta}{2}, \nonumber \\ 
	\Lambda_{23}^{(3,\,3)}&=&{\cal A}(y,\alpha,\beta)\,\sin \alpha\, \sin\beta\, \cos\frac{\beta}{2}=\Lambda_{32}^{(3,\,3)}, \ \ 
	\Lambda_{33}^{(3,\,3)}=\frac{{\cal A}(y,\alpha,\beta)}{2}\,\left[2+y^2+4\,y \cos \alpha\, \cos^3 \frac{\beta}{2}+y^2\, \cos 2\beta\right], \nonumber    	
\end{eqnarray} 
where we have denoted
\begin{equation}
{\cal A}(y,\alpha,\beta)=\frac{1}{1+y^2+2\,y\,\cos \alpha 
\cos^3 \frac{\beta}{2}}.
\end{equation} 
The real symmetric $4\times 4$ matrix  $\Omega^{(3,\,3)}$ constructed from $\Lambda^{(3,\,3)}$  is given by 
\begin{eqnarray}
	\label{o33}
	\Omega^{(3,\,3)}&=&\Lambda^{(3,\,3)}\,G\, \left(\Lambda^{(3,\,3)}\right)^T\nonumber \\ 
	&=&{\cal B}(y,\alpha,\beta)\ba{cccc} 3+\cos \beta & \sin\beta & 0 & 1+\cos \beta  \\ 
	\sin \beta & -(1+\cos \beta) & 0 & \sin \beta  \\ 0 & 0 & -(1+\cos \beta) & 0 \\ 
	1+\cos \beta & \sin\beta & 0 & -(1-\cos \beta) \ea 
\end{eqnarray} 
where 
\be
\label{calb}
{\cal B}(y,\alpha,\beta)=\frac{y^2 (1-\cos \beta)^2}{2\left(1+y^2+2\,y\cos \alpha 
	\cos^3 \frac{\beta}{2} \right)^{2}}.
\ee
Eigenvalues  of the matrix  $G\,\Omega^{(3,\,3)}$  are  $2$-fold degenerate and are given by  
\begin{eqnarray}
	\label{ev33}
	\lambda_0&=&2\,{\cal B}(y,\alpha,\beta)\nonumber \\ 
	&=&\frac{y^2 (1-\cos \beta)^2}{\left(1+y^2+2\,y\cos \alpha 
		\cos^3 \frac{\beta}{2} \right)^{2}},  \\ 
	     \lambda_1&=&{\cal B}(y,\alpha,\beta)\,\,(1+\cos \beta)\nonumber \\ 
	     &=&\frac{y^2 \sin^2\beta\, (1-\cos \beta)}{2\left(1+y^2+2\,y\cos \alpha 
		\cos^3 \frac{\beta}{2} \right)^{2}}.  
\end{eqnarray}
Except in the case of $\beta=\pi$, the eigenvector $X^{(3,3)}$ of the matrix $G\,\Omega^{(3,\,3)}$ associated with the highest eigenvalue $\lambda_0$ satisfy the condition $\left(X^{(3,3)}\right)^T\,G\,X^{(3,3)}=0$, confirming that the matrix $\Omega^{(3,\,3)}$ can be reduced to the form (\ref{o2c}) under SLOCC transformation.  The Lorentz transformation matrix  $L^{(3\,3)}_{A_2}$, which takes $\Omega^{(3,\,3)}$ to its canonical form $\widetilde{\Omega}^{(3,\,3)}$ is then explicitly constructed following the detailed approach of Ref.~\cite{supra}:  
\begin{eqnarray}
	\label{L33c}
L^{(3\,3)}_{A_2}=\ba{cccc} \frac{3+\cos \beta}{2\sin \beta}
	& \frac{1+\cos \beta}{\sin \beta} & 0 &  \frac{1+\cos \beta}{2\sin \beta}  \\ 
	-1 & -1 & 0 & -1 \\ 
	0 & 0 & -1 & 0 \\ 
	- \frac{1+3\cos \beta}{2\sin \beta}  & - \frac{1+\cos \beta}{\sin \beta}  & 0 &  \frac{1-\cos \beta}{\sin \beta} \ea, \ \ \beta\neq \pi.
\end{eqnarray} 
We have   
\begin{eqnarray}
	\label{omega33c}
	\widetilde{\Omega}^{(3,\,3)}&=&L^{(3\,3)}_{A_2}\,  \Omega^{(3,\,3)}\, \left(L^{(3\,3)}_{A_2}\right)^T \nonumber \\ 
	&=&2\, {\cal B}(y,\alpha,\beta)\, \ba{cccc} 1 & 0 & 0 & 0 \\
	0 & -\cos^2\frac{\beta}{2}  & 0 & 0 \\
	0 & 0  & -\cos^2\frac{\beta}{2} & 0 \\
	0 & 0 & 0 & -1\ea,\ \ \beta\neq \pi.
\end{eqnarray}
Note that 
\begin{equation}
	\label{phi330}
	\phi_0=\left(L^{(3\,3)}_{A_2}\,  \Omega^{(3,\,3)}\, \left(L^{(3\,3)}_{A_2}\right)^T\right)_{00}=2\,{\cal B}(y,\alpha,\beta)=\lambda_0
\end{equation}
and hence the canonical form $\widetilde{\Omega}^{(3,\,3)}$ is  diagonal (see (\ref{o2c})). 
\begin{figure}[h]
	\begin{center}
		\includegraphics*[width=3.5in,keepaspectratio]{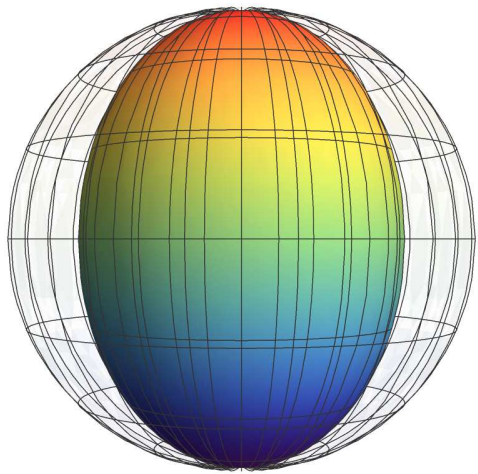}
		\caption{(Colour online)  Prolate spheroid centered at the origin of the Bloch sphere representing  the SLOCC canonical form of the two-qubit density matrix $\widetilde{\rho}^{\,(2-{\rm qubit})}_{3,\,3}$ given in (\ref{rr33c}). Semiaxes lengths of the spheroid are given by $\left(\cos\frac{\beta}{2}, \cos\frac{\beta}{2}, 1\right)$. Here we have chosen  $\beta=\pi/2$. }
	\end{center}
\end{figure}
Consequently the Lorentz canonical form 
$\widetilde{\Lambda}^{(3,\,3)}$ (see (\ref{lambda2c})) too is diagonal  
\begin{eqnarray}
	\label{l33c}
	\widetilde{\Lambda}^{(3,\,3)}&=&\ba{cccc} 1 & 0 & 0 & 0 \\
	0 & \cos\frac{\beta}{2}  & 0 & 0 \\
	0 & 0  & -\cos\frac{\beta}{2} & 0 \\
	0 & 0 & 0 & 1\ea,\ \ \beta\neq \pi 
\end{eqnarray}
and SLOCC canonical form $\widetilde{\rho}^{\,(2-{\rm qubit})}_{3,\,3}$ of the reduced two-qubit density matrix obtained by tracing one of the qubits of the pure symmetric three-qubit state $\vert\Psi^{ABC}_{3,\,3}\rangle$ (see (\ref{3meyer})) reduces to the following simple form :    
\begin{eqnarray}
	\label{rr33c}
	\widetilde{\rho}^{\,(2-{\rm qubit})}_{3,\,3}&=& \frac{1}{4}\, \left[ \, \sigma_0\otimes \sigma_0  + \cos\frac{\beta}{2} \,(\sigma_1\otimes\sigma_1 - \sigma_2\otimes\sigma_2) + \sigma_3\otimes\sigma_3\right]. 
\end{eqnarray} 
The Lorentz canonical form (\ref{l33c}) is represented by a prolate spheroid centered at the origin $(0,0,0)$ of  the Bloch sphere, with longest semiaxis $a_0$ along the $z$-direction (symmetry axis of the spheroid)  equal to  1 and  radius of the circular crosssection, perpendicular to the symmetry axis, given by $a_1=\cos\frac{\beta}{2}$. The steering spheroid associated with  the set of all two-qubit states, which lie on the SLOCC orbit of  $\widetilde{\rho}^{\,(2-{\rm qubit})}_{3,\,3}$ (see (\ref{rr33c})), is depicted in Fig.~2 (for the value $\beta=\pi/2$ of the state parameter).  

When the parameters $\beta=\pi,\ \alpha=0,\, y=1 $, the state $\vert\Psi^{ABC}_{3,\,3}\rangle$ (see (\ref{3meyer})) corresponds to the three-qubit GHZ state  (see (\ref{GHZ})). 
It can be readily seen from (\ref{o33}) that the matrix 	$\Omega^{(3,\,3)}$ is diagonal: 
\begin{eqnarray*}
\Omega^{(3,\,3)}={\rm diag}\left(1,\, 0,\, 0\, -1\right).
\end{eqnarray*}
From (\ref{lambda1c}) it is clear that the Lorentz canonical form of the matrix $\Lambda^{(3,\,3)}$ is given by   
\begin{eqnarray*}
	\widetilde{\Lambda}^{(3,\,3)}={\rm diag}\left(1,\, 0,\, 0,\,   1\right)
\end{eqnarray*}
This corresponds geometrically to a line joining the north and the south poles of the Bloch sphere.

\section{Volume monogamy relations and  entanglement} 

 Monogamy relations restrict shareability of quantum correlations in a multipartite state. They find potential applications in ensuring security in quantum key distribution~\cite{Tehral,Paw}.   Milne {\em et. al.}~\cite{MilneNJP2014} introduced a geometrically intuitive  monogamy relation for the volumes	of the steering ellipsoids representing  the two-qubit subsystems of a pure three-qubit state, which is stronger than
	the well-known  Coffman-Kundu-Wootters monogamy relation~\cite{CKW}. 	
	In this section we explore  how volume monogamy relation~\cite{MilneNJP2014} imposes limits on the volumes of the quantum steering ellipsoids representing the two-qubit subsystems   $\rho^{(2-{\rm qubit})}_{3,\,2}={\rm Tr}_{A}\,\vert\Psi^{ABC}_{3,\,2}\rangle\langle \Psi^{ABC}_{3,\,2}\vert$, $\rho^{(2-{\rm qubit})}_{3,\,3}={\rm Tr}_{A}\,\vert\Psi^{ABC}_{3,\,3}\rangle\langle \Psi^{ABC}_{3,\,3}\vert$ (see (\ref{rrho32}), (\ref{rrho33})) of the  SLOCC inequivalent families ${\cal D}_{3,2}$, ${\cal D}_{3,3}$ of pure symmetric three-qubit states. We also study the upper bounds imposed on the concurrence of reduced two-qubit states of pure three-qubit symmetric states and their SLOCC canonical forms discussed in Sec.~5.     

The volume of the quantum steering ellipsoid ${\mathcal E}_{A\vert\,B}=\left\{\mathbf{p}=\frac{\mathbf{s}\,+\,T^{T}\,\mathbf{e}}{1+\,\mathbf{r}\cdot\,\mathbf{e}}\right\}$ of a  two-qubit state $\rho_{AB}$  (see (\ref{rho2q})) is given by~\cite{jevtic2014} 
\begin{equation}
\label{new1}
	V_{B\vert A}=\left(\frac{4\pi}{3}\right)\, \frac{\vert\det \Lambda \vert}{(1-r^2)^2}, 
\end{equation} 
where  $r^2=\mathbf{r}\cdot\mathbf{r}=r_1^2+r_2^2+r_3^2$ (see \ref{ri}).  As the steering ellipsoid is constrained to lie within the Bloch sphere  $V_{B\vert A}\leq V_{\rm unit}=(4\pi/3).$ 

Volume of a steering ellipsoid captures several interesting  quantum correlation features~\cite{jevtic2014,MilneNJP2014,MilnePRA2016}, which we list below: 
\begin{itemize}
\item[(a)] Volumes $V_{B\vert A}$, $V_{A\vert B}$  of the quantum steering ellipsoids ${\cal E}_{B\vert A}$ and ${\cal E}_{A\vert B}$ are related by~\cite{jevtic2014} 
\begin{equation}
	\frac{ V_{A\vert B}}{(1-r^2)^2}= \frac{V_{B\vert A}}{(1-s^2)^2}. 
\end{equation}
	\item[(b)] An upper bound for the concurrence $C(\rho_{AB})$ is placed in terms of the {\em obesity} ${\cal O}(\rho_{AB})=\vert \det\Lambda\vert^{1/4}$ of the quantum steering ellipsoid~\cite{MilneNJP2014}:  
	 \begin{equation}
	 	\label{c&v}
	C(\rho_{AB})\leq {\cal O}(\rho_{AB}) =\vert \det\Lambda\vert^{1/4}.
	 \end{equation}
 Furthermore, if $\rho_{AB}\longrightarrow\widetilde{\rho}_{AB}=(A\otimes B)\rho_{AB}\, (A^\dag\otimes B^\dag)/({\rm Tr}(A^\dag\,A\otimes B^\dag B)\rho_{AB}]$, $A,B\in SL(2,C)$   it follows that~\cite{MilneNJP2014} 
 \begin{equation}
 	\label{cvratio}
 \frac{{\cal O}(\rho_{AB})}{C(\rho_{AB})}=\frac{{\cal O}(\widetilde{\rho}_{AB})}{C(\widetilde{\rho}_{AB})}
 	\end{equation}
 
\item[(c)] While steering ellipsoids of a two-qubit state lie inside the Bloch sphere, not all ellipsoids
contained within the Bloch sphere correspond to  {\em bonafide}  states.  Physical states impose limits on the volume of steering ellipsoid for a given centre. The extremal steering ellipsoids are those 
with  largest volume for a fixed centre $\mathbf{r}_0$ that can fit inside the Bloch sphere. Oblate spheroid of volume $V=(4\pi/3)(1-r_0)^2$, with major semiaxes $a=\sqrt{1-r_0}=b$ and radially oriented minor semiaxis $c=(1-r_0)$, $r_0\neq 1$ with centre $\mathbf{r}_0=(0, 0, r_0)$ correspond to extremal ellipsoid of entangled two-qubit state~\cite{MilneNJP2014}.       
 \item[(d)]  The two-qubit Werner state on the separable-entangled boundary has volume of the steering ellipsoid $V^{\rm Werner}_{A\vert B}=V^{\rm Werner}_{B\vert A}=(4\pi/81)=V_{\star}$, which happens to be the volume of largest possible tetrahedron that can be inscribed inside Bloch sphere~\cite{jevtic2014}. Separable two-qubit systems do not exhibit volume of their steering ellipsoids  larger than $V_\star=(4\pi/81)$. Thus,  $V>V_\star$ provides an intuitive geometric condition for entanglement in two-qubit states~\cite{jevtic2014}.   
\item Let Alice, Bob and Charlie share a pure three qubit state. Suppose that Bob performs all possible local measurements on his qubit so as to steer the states possessed by Alice and Charlie. The volumes $V_{A\vert B}, \ V_{C\vert B}$ of  the  resulting steering ellipsoids obey monogamy relation given by~\cite{jevtic2014}  
\begin{equation}
	\label{vm}
	\sqrt{V_{A\vert B}} + \sqrt{V_{C\vert B}} \leq \sqrt{\frac{4\pi}{3}}. 
\end{equation} 
\end{itemize}

\subsection{Volume monogamy relations governing the states $\vert\Psi^{ABC}_{3,\,2}\rangle$ of the SLOCC class ${\cal D}_{3,2}$ }
Let Alice, Bob, and Charlie share a pure entangled three-qubit symmetric state $\vert\Psi^{ABC}_{3,\,2}\rangle$ (see (\ref{2meyer})) of the SLOCC class ${\cal D}_{3,2}.$ Bob's local measurements generate identical steering ellipsoids ${\cal E}_{A\vert B}={\cal E}_{C\vert B}={\cal E}_{3,2}$ of Alice and Charlie because the reduced two-qubit density operators drawn from the pure three qubit state $\vert\Psi^{ABC}_{3,\,2}\rangle$ are all  same i.e.,  $\rho_{AB}=\rho_{BC}=\rho_{AC}=\rho^{(2-{\rm qubit})}_{3,\,2}$ (see (\ref{rrho32})). Evidently the volumes $V_{A\vert B}$, $V_{C\vert B}$ of the steering ellipsoids are equal. Denoting  $V_{3,2}=V_{A\vert B}= V_{C\vert B}$, the monogamy relation (\ref{vm}) takes the form, 
\begin{equation}
	\label{vm32}
	\sqrt{\frac{3\,	V_{3,2}}{\pi}}\leq 1 
	\end{equation}
for pure symmetric three-qubit states  $\vert\Psi^{ABC}_{3,\,2}\rangle$ (see (\ref{2meyer})) of the SLOCC class ${\cal D}_{3,2}$.
 
From (\ref{new1}), we have
\begin{equation} 
	\label{v32}
	V_{3,2}=\left(\frac{4\pi}{3}\right)\, \frac{\vert\det \Lambda^{(3,2)}\vert}{(1-r^2)^2}, 
\end{equation} 
where $\Lambda^{(3,2)}$ is given in (\ref{lambda32}) and 
\begin{equation}
	\label{r32}
	r_1=\frac{\sin\beta}{2+\cos\beta},\ r_2=0, r_3=\frac{5+4 \cos\beta}{3(2+\cos\beta)} 
\end{equation}
Under suitable SLOCC the state $\rho^{(2-{\rm qubit})}_{3,\,2}$ and the associated real matrix $\widetilde{\Lambda}^{(3,\,2)}$ get transformed to their respective canonical forms (see (\ref{l32c}), (\ref{rr32c}))
\begin{eqnarray*}
\widetilde{\rho}^{(2-{\rm qubit})}_{3,\,2}&=&\frac{(A \otimes B)\, \rho^{(2-{\rm qubit})}_{3,\,2}\, (A^\dag\otimes B^\dag)}
{{\rm Tr}\left[\rho^{(2-{\rm qubit})}_{3,\,2}\, (A^\dag\, A\otimes B^\dag\, B)\right]} \\ 
&=& \frac{1}{4}\, \left[ \, \sigma_0\otimes \sigma_0 + \frac{1}{2}\,\sigma_3\otimes \sigma_0 + \frac{1}{\sqrt{2}} \,(\sigma_1\otimes\sigma_1 - \sigma_2\otimes\sigma_2) +\,\frac{1}{2}\, \sigma_3\otimes\sigma_3\right]
\end{eqnarray*} 
and 
\begin{eqnarray*}
\Lambda^{(3,2)}\longrightarrow \widetilde{\Lambda}^{(3,2)}&=&\frac{L_A\,\Lambda\, L^T_B}{\left(L_A\,\Lambda\, L^T_B\right)_{00}}=
\left(\begin{array}{cccc}1 & 0 &0& 0  \\ 
	0& \frac{1}{\sqrt{2}} & 0  &  0\\ 
	0 & 0 &  -\frac{1}{\sqrt{2}}  & 0 \\ 
	\frac{1}{2}  &  0 & 0 &  \frac{1}{2}\end{array} \right).  
\end{eqnarray*} 
where (see (\ref{phi0}), (\ref{lphi}), \ref{ev32}))
\begin{equation}
	\label{phi032}
\left(L_A\,\Lambda\, L^T_B\right)_{00}=\sqrt{\phi_0}=\sqrt{2}\left[\frac{1-\cos\beta}{3(2+\cos\beta)}\right]. 
\end{equation}
Using the property $\det L_A=\det L_B=1$ of orthochronous proper Lorentz transformations~\cite{KNS} and substituting $\vert\det\widetilde{\Lambda}_{3,2}\vert=1/4$,  we obtain 
\begin{eqnarray}
	\label{det32}
\vert\det\widetilde{\Lambda}_{3,2}\vert=\frac{1}{4}&=&\vert\det L_A\vert\, \vert\det L_B\vert   \left\vert\det\left(\frac{\Lambda^{(3,2)}}{\sqrt{\phi_0}}\right)\right\vert  
=\frac{ \vert\det\,\Lambda^{(3,2)}\vert}{\phi_0^2}.  
\end{eqnarray}
Substituting (\ref{r32}), (\ref{phi032}) and (\ref{det32}) in (\ref{v32}), we obtain a simple form for the volume of the corresponding steering ellipsoid associated with  the two-qubit state $\rho^{(2-{\rm qubit})}_{3,\,2}$: 
\begin{eqnarray}
	V_{3,2}= \left(\frac{4\pi}{3}\right)\, \frac{\phi_0^2}{4\,(1-r^2)^2}=\frac{\pi}{3}. 
	\end{eqnarray} 
Evidently the volume monogamy relation (\ref{vm32}) is saturated by the family of three-qubit pure entangled symmetric states $\vert\Psi^{ABC}_{3,\,2}\rangle$ belonging to the SLOCC class ${\cal D}_{3,2}.$   Milne et al. have shown  that the volume monogamy relation (\ref{vm}) is saturated if and only if the steering ellipsoid has maximum volume for a given center. Thus, the reduced two-qubit states  of  the ${\cal D}_{3,2}$ class of three-qubit pure states have maximal volume steering ellipsoids. Coincidentally, this feature is also reflected in the canoncial steering ellipsoid  of  $\widetilde{\rho}^{\,(2-{\rm qubit})}_{3,\,2}$ (see (\ref{rr32c})), which has its center at $\mathbf{r}_0=(0,0,1/2)$, semi-major axes $a=b=\frac{1}{\sqrt{2}}$ and semi-minor axis $c=1/2$.  

\subsection{Volume monogamy relations governing the states $\vert\Psi^{ABC}_{3,\,3}\rangle$ of the SLOCC class ${\cal D}_{3,3}$ }
Suppose that Alice, Bob, and Charlie possess a qubit each of the permutation symmetric three-qubit pure entangled state $\vert\Psi^{ABC}_{3,\,3}\rangle$ (see (\ref{3meyer})). As the reduced two-qubit states are all identical i.e., $\rho_{AB}=\rho_{BC}=\rho_{AC}=\rho^{(2-{\rm qubit})}_{3,\,3}$ (see (\ref{rrho33}))  the steering ellipsoids  of Alice and Charlie generated by Bob's local measurements are also identical i.e., 
${\cal E}_{B\vert A}={\cal E}_{C\vert A}={\cal E}_{3,3}$. We thus have $V_{A\vert B}= V_{C\vert B}=V_{3,3}$ and the volume monogamy relation (\ref{vm})  for  pure symmetric three-qubit states  $\vert\Psi^{ABC}_{3,\,3}\rangle$ (see (\ref{3meyer})) of the SLOCC class ${\cal D}_{3,3}$ can be expressed as 
\begin{equation}
	\label{vm33}
	\sqrt{\frac{3\,V_{3,3}}{\pi}}\leq 1. 
\end{equation}

We evaluate the volume of the steering ellipsoid corresponding to the state $\rho^{(2-{\rm qubit})}_{3,\,3}$ (see (\ref{rrho33})))
\begin{equation}
	\label{v33}
	V_{3,3}=\left(\frac{4\pi}{3}\right)\, \frac{\vert\det\Lambda^{(3,3)}\vert}{(1-r^2)^2}
	\end{equation}
where (see (\ref{l33e})) 
\begin{eqnarray}
	\label{r33}
	r_1&=&\frac{y\,\sin\beta\,\left(y+\cos \alpha\, \cos \frac{\beta}{2}\right)}{1+y^2+2\,y\,\cos \alpha \cos^3\frac{\beta}{2}},\ \ r_2=\frac{y\, \, \sin \alpha\,\sin\beta\, \cos\frac{\beta}{2}}{1+y^2+2\,y\,\cos \alpha \cos^3\frac{\beta}{2}} \nonumber \\
	&& \hskip 1in 	r_3= 	\frac{1+2\,y\, \cos \alpha\, \cos^3\frac{\beta}{2}+y^2\,\cos \beta}{1+y^2+2\,y\,\cos \alpha \cos^3\frac{\beta}{2}} 
		\end{eqnarray} 
by expressing it in terms of the volume $\widetilde{V}_{3,3}=(4\pi/3)\, \cos^2(\beta/2)$ of the canonical steering ellipsoid specified by its radially aligned longest semi-major axis $c=1$ and semi-minor axes $a=b=\cos(\beta/2)$ (see (\ref{l33c})).

Substituting $\vert\det\widetilde{\Lambda}_{3,3}\vert=\cos^2(\beta/2)$ for the determinant of the Lorentz canonical form $\widetilde{\Lambda}_{3,3}$ given by (\ref{l33c}) and simplifying, we obtain 
\begin{eqnarray}
	\label{det33}
\cos^2\frac{\beta}{2}=\vert\det\widetilde{\Lambda}_{3,3}\vert &=& \left\vert\det\, \left(\frac{L_A\, \Lambda^{(3,3)}\, L_B^T}{\left(L_A\, \Lambda^{(3,3)}\, L_B^T\right)_{00}}\right)\right\vert 
=\phi_0^{-2}\, \left\vert\det\, \Lambda^{(3,3)} \right\vert, 
\end{eqnarray}
where (see (\ref{phi0}), (\ref{ev33}), (\ref{phi330})). 
\begin{equation}
	\label{phi33}
\phi_0=\frac{ y^2 (1-\cos\beta)^2}{(1+y^2+2\,y\cos \alpha	\cos^3\frac{\beta}{2})^2}
\end{equation} 
Thus,
\begin{equation}
	\label{detlambda33}
\left\vert\det\, \Lambda^{(3,3)} \right\vert=\phi_0^2 \cos^2\frac{\beta}{2}.
\end{equation}
After simplification (substituting (\ref{r33}), (\ref{detlambda33}) in (\ref{v33}) and using (\ref{phi33})) we obtain  
\begin{eqnarray}
	\label{v33s}
\frac{3\, V_{3,3}}{\pi}&=& \frac{4\,\phi_0^2 \cos^2\frac{\beta}{2}}{(1-r^2)^2}	
       =\frac{4\,\cos^2\frac{\beta}{2}}{(1+ \cos^2\frac{\beta}{2})^2}. 
\end{eqnarray}
Thus, the volume monogamy relation (\ref{vm33}) gets reduced to the simple form 
\begin{equation}
	\label{v33s}
 \frac{2\,\cos\frac{\beta}{2}}{(1+ \cos^2\frac{\beta}{2})}	\leq 1
\end{equation} 
which is  evidently obeyed by the set of all three-qubit pure symmetric states $\vert\Psi^{ABC}_{3,\,3}\rangle$ (See Fig.~3).  
\begin{figure}[h]
	\begin{center}
		\includegraphics*[width=3.5in,keepaspectratio]{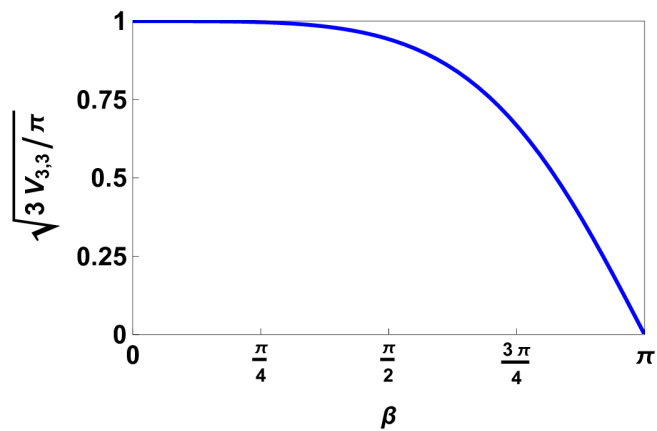}
		\caption{(Colour online) Left hand side $\sqrt{\frac{3\,V_{3,3}}{\pi}}=\frac{2\,\cos\frac{\beta}{2}}{(1+ \cos^2\frac{\beta}{2})}$ (see (\ref{v33s}))  of the volume monogamy relation (\ref{vm33}) governing the set of all three-qubit pure symmetric states $\vert\Psi^{ABC}_{3,\,3}\rangle$ belonging to the SLOCC family ${\cal D}_{3,3}$. } 
	\end{center}
\end{figure}  
 
\subsection{Connection between obesity of the steering ellipsoid and concurrence}
The concurrence of a two qubit state $\rho_{AB}$ is given by~\cite{Wootters} 
\begin{equation}
C(\rho_{AB})={\rm max} (0,\mu_1-\mu_2-\mu_3-\mu_4),
\end{equation}
where $\mu_1,\mu_2,\mu_3,\mu_4$ are the square-roots of the eigenvalues, in decreasing order,  of the matrix \break $\rho_{AB}\,(\sigma_2\otimes\sigma_2)\, \rho^*_{AB}\, (\sigma_2\otimes\sigma_2).$

We  evaluate the concurrence for the SLOCC canonical state $\widetilde{\rho}^{(2-{\rm qubit})}_{3,\,2}$ given by (\ref{rr32c}) to obtain 
\begin{equation}
	C(\widetilde{\rho}^{(2-{\rm qubit})}_{3,\,2})=\frac{1}{\sqrt{2}}.	
\end{equation}
The obesity of the canonical steering ellipsoide of $\widetilde{\rho}^{(2-{\rm qubit})}_{3,\,2}$ is given by (see (\ref{l32c}))
\begin{eqnarray}
	\label{obc32}
	{\cal O}(\widetilde{\rho}^{(2-{\rm qubit})}_{3,\,2})&=&\vert \det \widetilde{\Lambda}^{(3,2)}\vert^{1/4}=\frac{1}{\sqrt{2}}.
\end{eqnarray}
We also obtain the obesity 	${\cal O}(\rho^{(2-{\rm qubit})}_{3,\,2})$ using (\ref{phi032}), (\ref{det32}): 
\begin{eqnarray}
	\label{ob32}
	{\cal O}(\rho^{(2-{\rm qubit})}_{3,\,2})&=&\vert \det \Lambda^{(3,2)}\vert^{1/4}=\left[\frac{1-\cos\beta}{3(2+\cos\beta)}\right].
\end{eqnarray}
Based on  the relation (\ref{cvratio}) we compute the concurrence for the two-qubit state $\rho^{(2-{\rm qubit})}_{3,\,2}$ as 
\begin{equation}
	\label{c32}
C(\rho^{(2-{\rm qubit})}_{3,\,2})=\frac{1-\cos\beta}{3(2+\cos\beta)}
\end{equation}
which matches perfectly with the expression  derived  previously in \cite{meyer}. 

It is straightforward to check (see \ref{ob32}), (\ref{c32})) that the inequality (\ref{c&v}) gets saturated in $\rho^{(2-{\rm qubit})}_{3,\,2}$, in accordance with the result that the maximum volume steering ellipsoids maximize concurrence~\cite{MilneNJP2014}.
For the Bell-diagonal two-qubit state $\widetilde{\rho}^{(2-{\rm qubit})}_{3,\,3}$ (see (\ref{rr33c}) the concurrence is readily evaluated to be 
\begin{equation}
C(\widetilde{\rho}^{(2-{\rm qubit})}_{3,\,3})=\cos\frac{\beta}{2} 
\end{equation}
and the obesity of the canonical steering ellipsoid of $\widetilde{\rho}^{(2-{\rm qubit})}_{3,\,3}$ (see (\ref{l33c})) is given by 
\begin{equation}
{\cal O}(\widetilde{\rho}^{(2-{\rm qubit})}_{3,\,3})= \left\vert\det\, \widetilde{\Lambda}_{3,3} \right\vert^{1/4} =\sqrt{\cos\frac{\beta}{2}}
\end{equation}
Using (\ref{det33}), (\ref{phi33})  we compute the obesity  
\begin{equation}
	\label{o33}
	{\cal O}(\rho^{(2-{\rm qubit})}_{3,\,3})= \left\vert\det\, \Lambda^{(3,3)} \right\vert^{1/4} = \frac{ y\, \, (1-\cos\beta)\,\sqrt{\cos\frac{\beta}{2}}}{(1+y^2+2\,y\cos \alpha	\cos^3(\beta/2))}
\end{equation}
and obtain the  concurrence for $\rho^{(2-{\rm qubit})}_{3,\,3}$ based on (\ref{cvratio})
\begin{equation}
	\label{c33}
	C(\rho^{(2-{\rm qubit})}_{3,\,3})=\frac{ y\, \sin\beta\, \sin\frac{\beta}{2}}{(1+y^2+2\,y\cos \alpha	\cos^3(\beta/2))},
\end{equation}
which is clearly in agreement with the expression for concurrence given by Meill and Meyer~\cite{meyer}.   
It is easy to see that the limiting condition (\ref{c&v}))  imposed  on the concurrence (\ref{c33}) by the obesity (\ref{o33}) of the steering ellipsoid reduces to a simple trignometric relation  $\sqrt{\cos\frac{\beta}{2}}\leq \cos\frac{\beta}{2}.$  
\section{Summary}
We have shown that SLOCC inequivalent classes of pure entangled three-qubit  symmetric states come with {\em distinct} geometric visualization in terms of  canonical steering spheroid inscribed within the Bloch sphere.   We have evaluated explicit analytical structure for the SLOCC canonical forms associated with the reduced two-qubit states -- obtained by tracing out one of the qubits of entangled three-qubit pure symmetric states. Our analysis yields a clear geometric insight: Reduced two-qubit states of a pure permutation symmetric three-qubit state $\vert\Psi^{ABC}_{3,\,2}\rangle$, which is constructed by symmetrizing 2 distinct qubits -- is represented by an extremal steering spheroid centered at $(0,0,1/2)$ inside  the Bloch sphere, with fixed semiaxes lengths $(1/\sqrt{2},\, 1/\sqrt{2},\, 1/2)$. On the other hand, the two-qubit density matrices drawn from the three-qubit state $\vert\Psi^{ABC}_{3,\,3}\rangle$ constituted by permutation of 3 distinct qubits is geometrically viewed as  a prolate spheroid centered at the origin of the Bloch sphere -- with  length of the semi-major axis along the symmetry direction of the spheroid  equal to 1. The canonical steering ellipsoids are shown to offer insightful information on different aspects of quantum correlations.  We have illustrated volume monogamy relations  obeyed  by the three-qubit pure symmetric states belonging to  the SLOCC inequivalent classes ${\cal D}_{3,2}$, ${\cal D}_{3,3}$. Furthermore, connection between  the obesity of the steering ellipsoid and concurrence of the two-qubit subsystems of  three-qubit pure symmetric states is elucidated.

\section*{Acknowledgements} AK acknowledges  University Grants Commission (RGNF) for financial support; Sudha, ARU and IR are supported by the Department of Science and Technology, India
(Project No. DST/ICPS/QUST/2018/107);  HSK acknowledges the support of NCN through SHENG Grant No. 2018/30/Q/ST2/00625.  

\section*{Data avialability statement} Data sharing not applicable to this article as no datasets were generated or analysed during the current study.

\end{document}